\documentclass[%reprint,
  aps,
  pra,
  reprint,
  amsmath,
  amssymb,longbibliography
]{revtex4-2}

\usepackage[colorlinks=true,urlcolor=blue,citecolor=blue,allcolors=blue]{hyperref}

\usepackage{graphicx}% Include figure files
\usepackage{dcolumn}% Align table columns on decimal point
\usepackage{bm}% bold math
\usepackage{braket}
\usepackage{xcolor}
\usepackage[dvipsnames]{xcolor}

\begin{document}

\newcommand{\rev}[1]{{\color{black}#1}}
\newcommand{\todo}[1]{{\color{red}#1}}

\preprint{APS/123-QED}

\title{Refrigeration of a one-dimensional gas of microwave photons}

\author{Lukas Schamri\ss}
 \email{lukas.schamriss@tum.de}

\author{Louis Garbe}%
 \author{Peter Rabl}
 \affiliation{%
Technical University of Munich, TUM School of Natural Sciences, Physics Department, 85748 Garching, Germany
}%
\affiliation{%
Walther-Meißner-Institut, Bayerische Akademie der Wissenschaften, 85748 Garching, Germany
}%
\affiliation{%
Munich Center for Quantum Science and Technology (MCQST), 80799 Munich, Germany
}%

\date{\today}

\begin{abstract}
We discuss a conceptually simple scheme for cooling a one dimensional gas of microwave photons in a superconducting transmission line. By shunting one end of the transmission line by a nonlinear Josephson element, we show how a cooling mechanism can be engineered that  transfers photons from high- into low-frequency modes, while preserving their total number. We evaluate the resulting nonequilibrium steady state of the photon gas, which  arises from a competition between this engineered cooling process and the natural, number non-conserving thermalization with the surrounding bath. Our analysis predicts that for realistic experimental parameters, this mechanism can be used to prepare photonic gases at sub-millikelvin temperatures, considerably below the typical base temperature of a dilution refrigerator. In addition, the system exhibits a new type of condensation transition that does not occur in the corresponding equilibrium scenario. As an outlook, we discuss potential applications of this cooling approach for quantum simulation schemes with interacting microwave photons.  

\end{abstract}

\maketitle

\section{\label{sec:introduction}Introduction}
Over the past years, superconducting quantum circuits have emerged as one of the leading platforms for quantum technologies. While the main focus in this field is currently placed on digital quantum information processing and quantum simulations applications that can be run on already commercially available hardware, superconducting circuits also offer great opportunities for analog simulation \cite{Houck2012,Hartmann2016,Kjaergaard2020,Greenaway2024}. In particular, this platform benefits from the ability to implement very strong interactions \cite{Niemczyk2010,Forn-Diaz2017,Yoshihara2017}, 
to design custom lattice geometries \cite{Kollar2019} and long-range connectivity \cite{Zhang2023}, to engineer effective magnetic fields \cite{Roushan2017,Owens2018,Rosen2024,Wang2024}, and from many other control capabilities for both two-level systems and bosonic degrees of freedom. 
These experimental techniques provide a versatile set of tools to realize  a wide range of effective quantum many-body Hamiltonians, ranging from Bose- and Fermi-Hubbard-models \cite{GarciaRipoll2008,Koch2009,Leib2010,Reiner2016, Ma2019,Karamlou2024,Li2025,Du2025}
and fractional quantum Hall systems \cite{Wang2024} to lattice gauge theories \cite{Marcos2013,Marcos2014,Busnaina2025}
and gravity models \cite{Shi2023,Javed2024}.

Superconducting quantum circuits can be cooled to a base temperature of $T\approx 20$ mK, which is enough to initialize qubits or microwave resonators---with typical transition frequencies of about $\sim 3-5$ GHz---in their ground state. These temperatures are, however, still large compared to the relevant energy scales \rev{of about $\sim 10-100$ MHz} \cite{GarciaRipoll2008,Ma2019,Karamlou2024,Wang2024,Du2025} of the engineered many-body models of interest, which are often determined by much weaker Josephson nonlinearities or by higher-order perturbative processes. Furthermore, simply cooling the whole microwave circuit reduces the total number of excitations in the system and thus prevents one, for example, from investigating bosonic lattice models at fixed densities.  This is in stark contrast to quantum simulations schemes with cold atoms \cite{Gross2017}, and also differs from optical photons or polaritons, where efficient, phonon-induced thermalization processes \cite{Imamog¯lu1996,Deng2002,Kasprzak2006,Balili2007,Keeling2007,Deng2010,Klaers.2010} that approximately conserve the photon number are naturally available. Therefore, analog quantum simulation in superconducting circuits is usually constrained to quench dynamics or nonequilibrium steady-states in continuously driven systems, whereas---in the absence of a naturally occurring low-temperature reservoir---the preparation of ground and thermal equilibrium states remains an open problem.

To overcome this deficiency---at least in part---several schemes have been put forward to stabilize many-body states with finite photon numbers. These include, for example, the preparation of a Bose-Einstein condensate (BEC) of microwave photons \cite{Marcos2012} or a photonic Mott-insulator state \cite{Ma2019} in a coupled array of transmons. These schemes are, however, designed to stabilize a specific, previously known many-body state, and are thus not immediately applicable for the preparation of more general equilibrium states that appear in quantum simulations. Alternatively, active cooling schemes and the upconversion of low temperature fluctuations have been proposed as a way to mimic effective reservoirs for microwave photons, with tunable temperature and chemical potential \cite{Hafezi2015,Hacohen2015, Cao2025}. Similarly, iterative quantum algorithmic approaches to the cooling problem have been proposed~\cite{Polla2021,Marti2025}, which require only partial knowledge of the many-body energy spectrum. Although such approaches are  more generally applicable, a complete thermalization of photonic many-body systems that are coupled to such engineered reservoirs can only be guaranteed in certain idealized limits, which may not be reachable in experiments. Any residual decay, finite cooling rates and other imperfections lead to deviations from equilibrium, which are difficult to predict in advance for generic quantum many-body systems. \rev{In particular, the stability of the predicted stationary states under realistic experimental conditions is largely unexplored, but highly relevant for the application of engineered equilibrization schemes.}

In this paper, we propose and analyze a \rev{conceptually simple and experimentally feasible}  circuit QED setup, \rev{which is designed to prepare} a one-dimensional (1D) gas of microwave photons close to a  thermal equilibrium state with tunable temperature and a non-vanishing chemical potential. The purpose of this investigation is two-fold. Firstly, by focusing on the case of a non-interacting gas as the simplest instance of a photonic many-body system, we can accurately predict the resulting stationary distributions that arise from the competition between the engineered photon-number conserving cooling process and the thermalization with the surrounding electromagnetic environment. This allows us to characterize both equilibrium and the nonequilibrium aspects of the steady state, and to identify a meaningful, operatively-defined effective temperature to quantify cooling efficiencies in this setting. For an Ohmic environment, our analysis also reveals an unexpected feature of the steady state, namely that its total energy remains conserved \rev{when the coupling to the engineered bath is switched on}. In the thermodynamic limit, this constraint induces a sharp photon-condensation transition, which does not occur in an equivalent 1D system in equilibrium. 
\rev{This effect occurs for an arbitrarily weak residual coupling to the thermal electromagnetic environment and cannot be predicted from the analysis of the engineered cooling process by itself. This observation thus provides valuable insights for the stability and experimental robustness of other engineered many-body cooling schemes, for which an explicit evaluation of the steady-state is usually not possible.}

Secondly, our setup shows how multi-mode cooling processes can already be implemented using a single nonlinear Josephson coupler. Compared to related previous schemes \cite{Marcos2012,Hafezi2015}, this reduces considerably the experimental overhead, and makes first demonstrations of many-body cooling processes feasible in small-scale superconducting circuits with limited control. We provide a detailed microscopic derivation of the cooling master equation starting from a realistic circuit model, and show that even with this minimal setup, a gas of \rev{hundreds} of microwave photons can be cooled to sub-millikelvin temperatures. We also identify some of the main limitations of this approach, which arise primarily from residual Kerr nonlinearities and must be addressed to further improve such cooling schemes in future applications.        

\section{Model}

\begin{figure}
    \centering
    \includegraphics[width=\linewidth]{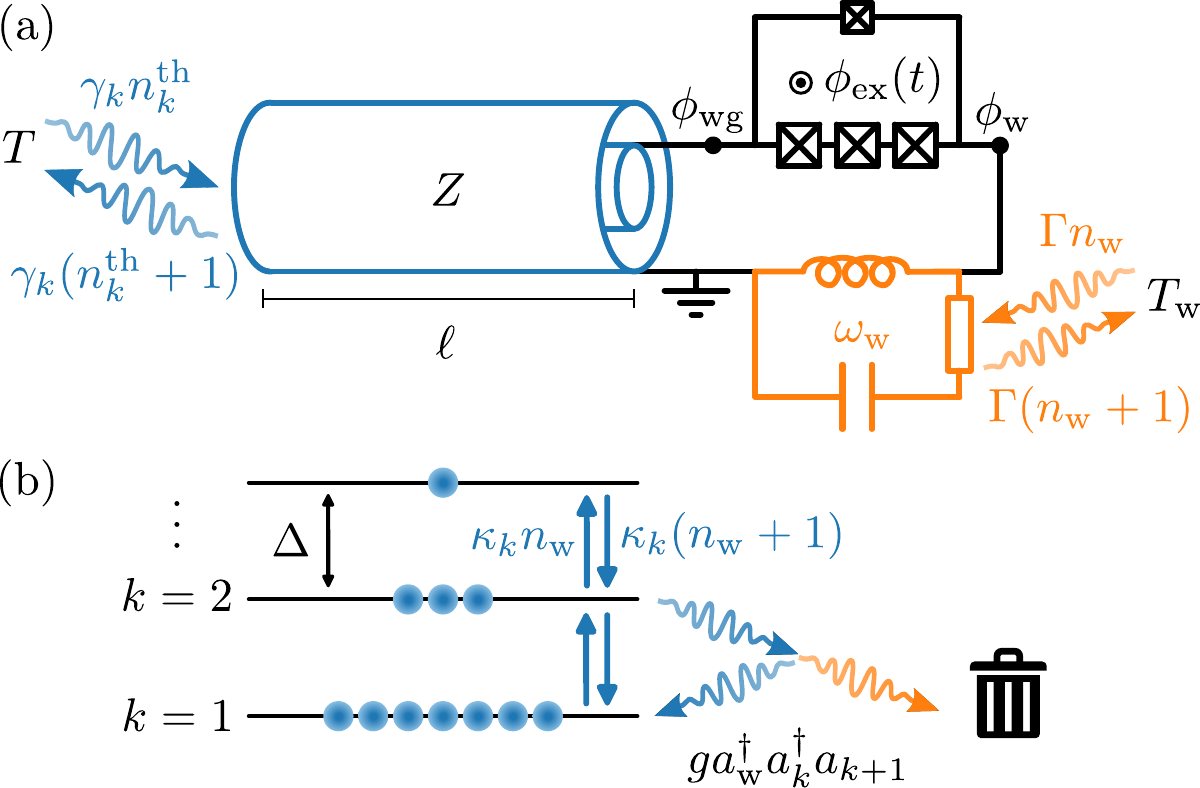}
    \caption{(a) Sketch of a refrigerator for microwave photons. A 1D transmission line resonator (blue) of length $\ell$ and mode spacing $\Delta$ is coupled to its thermal environment and to an engineered bath consisting of a lossy $LC$-resonator (orange), the waste mode. The coupling to the waste mode is mediated by an externally driven, nonlinear Josephson coupler to induce a resonant three-wave-mixing process.
    (b) Through this process, the photons in the waveguide can be scattered between neighboring modes, with the excess energy $\hbar \Delta$ being absorbed by the waste resonator and subsequently dissipated into the environment. This process is incoherent, but conserves the total number of photons in the 1D gas. The steady-state occupations $n_k$ are governed by the competition between this engineered cooling mechanism and the direct interaction of the waveguide with the environment.  }
    \label{fig:model}
\end{figure}

We consider a 1D gas of noninteracting microwave photons in a multimode transmission-line resonator of length $\ell$ [see Fig.~\ref{fig:model}(a)]. The resonator supports a discrete set of standing-wave modes with bosonic annihilation and creation operators $a_k$ and $a_k^\dag$ and frequencies 
\begin{equation}
\omega_k= \Delta  k,
\end{equation}
where $k=1,2,3,\dots,k_{\rm max}$ is the mode index running up to a cutoff value of $k_{\rm max}$ and $\Delta= \omega_{k+1}-\omega_k= v \pi /\ell$ is the mode spacing for a linear dispersion relation with group velocity $v$. Throughout this work, the cutoff $k_{\rm max}$ is set high enough to ensure that it has no impact on the results.

The photon gas is weakly coupled to the surrounding electromagnetic environment, which is held at a fixed temperature $T$. In the absence of any other couplings, the equilibrium occupation of the modes is therefore given by the usual Bose-Einstein distribution, 
\begin{equation}
n_k^{\rm th} = \langle a_k^\dag a_k\rangle_{\rm eq} = \frac{1}{e^{\beta\hbar \omega_k}-1},
\label{eq:BE_eq}
\end{equation}
where $\beta=1/(k_BT)$\rev{, $k_B$ is Boltzmann's constant and $\hbar$ is the reduced Planck constant}. By lowering the support temperature, the total energy of the system, ${E_{\rm ph}= \sum_k \hbar \omega_k \langle a^\dag_k a_k\rangle}$, can be reduced, but only at the expense of reducing the average total number of photons,  $N_{\rm ph} =\sum_k \langle a^\dag_k a_k\rangle$, as well.

\subsection{Reservoir engineering} 
To break this relation between energy and photon number, one end of the transmission line is coupled to an additional auxiliary resonator mode, the waste mode, with annihilation operator $a_\mathrm{w}$ and frequency $\omega_\mathrm{w}$. This coupling is implemented via a nonlinear Josephson element that induces a three-wave-mixing process, described by an Hamiltonian of the form:
\begin{equation}\label{eq:Hint}
H_\text{int}=\sum_k \frac{\hbar g}{\sqrt{k(k+1)}} \left(a_\mathrm{w}^\dagger a_k^\dagger a_{k+1} + a_\mathrm{w} a_k a^\dag_{k+1} \right),
\end{equation}
where $g$ is the characteristic, mode-independent coupling strength. The additional $k$-dependence arises from the zero-point fluctuations $\phi_k^0\sim 1/\sqrt{\omega_k}$ of the quantized flux variables. When the waste mode is tuned into resonance with the frequency difference between two neighboring modes, i.e., $\omega_\mathrm{w}=\Delta$, Eq.~\eqref{eq:Hint} describes a resonant conversion of a photon in mode $k+1$ into a photon in the lower frequency mode $k$ and a photon in the waste mode [see Fig.~\ref{fig:model}(b)]. The waste mode then quickly dissipates its energy into the environment. The net process is an incoherent redistribution of photons between neighboring waveguide modes~\cite{Lescanne2020,Nojiri2024} while preserving their total number, i.e., $\left[H_{\rm int},\sum_k a^\dag_k a_k\right]=0$.

In Sec.~\ref{sec::impl} below we discuss in more detail a specific circuit layout for implementing $H_{\rm int}$ with a time-periodic coupling, such that the external modulation frequency can be used to gap any frequency mismatch between $\omega_\mathrm{w}$ and $\Delta$. In this case, the resonant three-photon process in Eq.~\eqref{eq:Hint} can be realized independently of the bare energy scales and in the following we treat $\omega_\mathrm{w}$ and $\Delta$ as independently adjustable parameters. This external driving field
is the key ingredient that allows us to break thermal equilibrium.

\subsection{Cooling master equation}
To achieve cooling, we are interested in the limit where the waste mode quickly decays with a rate $\Gamma\gg g$. This means that this mode can be treated as an effective Markovian reservoir, and we can adiabatically eliminate its dynamics to obtain a master equation for the reduced state of the photon gas only. By also taking the bare losses of the waveguide modes into account, this master equation reads
\begin{equation}\label{eq:me}
\dot{\rho}=(\mathcal{L}_\text{cool}+\mathcal{L}_\text{th})\rho.
\end{equation}
Here, the first term describes the engineered, photon-number conserving dissipation. It is given by (see Sec.~\ref{sec::impl} for more details)
\begin{equation}
\begin{aligned}\label{eq:Lint}
\mathcal{L}_\text{cool} = \sum_{k=1}^{k_{\rm max}-1} \kappa_k\left\{ 
     (n_{\rm w}+1)\, \mathcal{D}\big[ a_k^\dagger a_{k+1} \big]  
  +n_{\rm w} \, \mathcal{D}\big[ a_{k+1}^\dagger a_k \big]  \right\},
\end{aligned}
\end{equation}
where $\mathcal{D}[c] \rho = c \rho c^\dagger - \frac{1}{2} \left\{ c^\dagger c, \rho \right\}$, $\kappa_k= g^2/[\Gamma k(k+1)]$ is the effective transition rate and $n_{\rm w}=\langle a_\mathrm{w}^\dagger a_\mathrm{w}\rangle_{\rm eq}=
[e^{\hbar \omega_\text{w}/(k_B T_{\rm w})}-1]^{-1}$ is the equilibrium occupation number of the waste mode. 
%
%\lou{If we consider observables instead of states, the evolution under a jump operator $c$ takes on the form:} 
%\lou{\begin{align}\label{eq:lindblad_operators}
%\langle \dot{O}\rangle=\text{Tr}[O\dot{\rho}]=\frac{1}{2}\langle c^\dag[O,c]\rangle+ \frac{1}{2}\langle[c^\dag,O] c\rangle
%\end{align}}
%
Note that for the sake of generality, we have introduced $T_{\rm w}$ as the relevant temperature of the reservoir to which the waste mode is coupled to. While for most results we assume $T=T_{\rm w}$, this temperature can be locally adjusted and used as an additional tuning knob to set the value of $n_{\rm w}$. When $n_{\rm w}\lesssim 1$, the second term in Eq.~\eqref{eq:Lint} is suppressed and $\mathcal{L}_\text{cool}$ describes an incoherent, but number-conserving transfer of photons from high- to low-frequency modes, i.e., cooling. 

This cooling process competes with the equilibration of the photons with the surrounding bath, which is described by $\mathcal{L}_\text{th}$. Assuming, for concreteness, an Ohmic coupling to the environment, the bare decay rates of each mode, $\gamma_k=\omega_k/Q$, are described by the same quality factor $Q$ and we obtain  
\begin{equation}\label{eq:Lloss}
    \mathcal{L}_\text{th}=\sum_{k=1}^{k_{\rm max}} \gamma_k\left\{\left(n^\text{th}_k+1\right)\mathcal{D}\big[a_k\big]+n^\text{th}_k\mathcal{D}\big[a_k^\dagger\big]\right\}
\end{equation}
with $\gamma_k = k\gamma $ and $\gamma=\Delta/Q$ denoting the decay rate of the fundamental mode. The action of $\mathcal{L}_\text{th}$, which arises from weak linear interactions with the environment, does not conserve the photon numbers and tends to relax the system back to its equilibrium state. 

Before we proceed with the analysis of Eq.~\eqref{eq:me}, let us remark that similar photon-number conserving master equations have been previously considered in related studies of photon condensation in the optical~\cite{Kirton2013,Kirton2015} and the microwave regime \cite{Marcos2012}, or in reservoir engineering schemes for cold atoms \cite{Diehl2008}. Similar dissipative processes as described by Eq.~\eqref{eq:Lint} also appear in the modeling of bosonic cascade lasers \cite{Liew2013,Kaliteevskii2013,Kavokin2016} and dissipative bosonic transport~\cite{Haga2021, Garbe2024, Minoguchi2025, Bernard2025}. While the common phenomenology of condensation and lasing effects are generally observed in such models, the detailed behavior of the system depends strongly on the spectral distributions of the modes and the engineered transition rates, and is analyzed here specifically for the microwave setup described above.

\section{Refrigeration of microwave photons}

\rev{To analyze the steady state of the system, we focus on the equations of motion for the average mode occupation numbers $n_k=\langle a_k^\dagger a_k^{\phantom{\dagger}}\rangle$, which are derived from the master equation in Eq.~\eqref{eq:me} and obey
%\lou{Combining Eq.\eqref{eq:Lint},\eqref{eq:lindblad_operators}, and \eqref{eq:Lloss}, we find the kinetic equations for these quantities:}
\begin{align}\label{eq:mf}
\dot{n}_k&=J_{k,k+1}-J_{k-1,k}-\gamma_k (n_k-n^\text{th}_k).
\end{align}
Here, $J_{k,k+1}$ are the currents of photons from mode $k+1$ to $k$, which are induced by the engineered cooling processes and given by
\begin{align}\label{eq:current}
J_{k,k+1}=\, &\kappa_k(1+n_{\rm w})\langle a_k a^\dag_k a^\dag_{k+1}a_{k+1}\rangle \nonumber \\
&- \kappa_k n_{\rm w}\langle a^\dag_k a_k a_{k+1} a^\dag_{k+1}\rangle.
\end{align}
Since the jump operators that appear in $\mathcal{L}_\text{cool}$ in Eq.~\eqref{eq:Lint} are nonlinear, these currents depend on higher moments of the number operators. To close the system of equations, we use a mean-field decoupling ${\langle a_k^\dagger a_k a_{k+1}^\dagger a_{k+1}\rangle\simeq n_kn_{k+1}}$ and approximate the currents by
\begin{align}\label{eq:mfcurrent}
J_{k,k+1}
\simeq \kappa_k\left[n_{\rm w}(n_{k+1}-n_k)+(n_k+1)n_{k+1}\right].
\end{align}
Note that in the limit $n_{\rm w}\rightarrow 0$ the currents become fully directional and photons only flow from high- to low-energy modes.}
%The second, non-linear term in this expression arises from the anti-commutation of $a_k,a^\dag_{k+1}$, and encodes the non-reciprocal properties of the transport.}
From \eqref{eq:mf}, the steady-state occupation numbers can be obtained by numerically solving the set of equations $\dot{n}_k=0$ \cite{mpmath,schamriss_2026}.

\subsection{Stationary photon-number distribution}

Within the mean-field approximation, the equations of motion given in Eq.~\eqref{eq:mf} and Eq.~\eqref{eq:mfcurrent} split into two parts that act as competing thermalization mechanisms for the photons in the waveguide. On the one hand, linear losses $\sim \mathcal{L}_{\rm th}\propto\gamma$ drive the system towards an equilibrium state at the temperature $T$ of the \rev{environment} [see Eq.~\eqref{eq:BE_eq}]. In the limit $g=0$, we will therefore recover $n_k|_{g=0}=n^{\text{th}}_k$. 
\rev{In contrast, for $\gamma=0$, the steady state is solely governed by the engineered cooling mechanism, and the resulting steady-state distribution is uniquely determined by the detailed balance conditions $J_{k,k+1}=0$. Therefore, it satisfies}
\begin{equation}
1+n_{k+1}^{-1}=e^{\hbar\omega_{\rm w}/(k_B T_{\rm w})}(1+n_k^{-1}).
\end{equation}
Importantly, although the process $\mathcal{L}_\text{cool}$ is fundamentally out-of-equilibrium, this relation still establishes a detailed balance between the modes $k$ and $k+1$. The resulting distributions can therefore be cast into the form
\begin{equation}\label{eq:efTh}
n_k|_{\gamma=0}=\frac{1}{e^{\beta_\mathrm{low}(\hbar\omega_k-\mu_\mathrm{eff})}-1},
\end{equation}
where $\beta_\mathrm{low}=1/(k_B T_\mathrm{low})$ and  $T_\mathrm{low}=T_{\rm w}\Delta/\omega_\text{w}$ is the engineered temperature. We refer to this distribution as the \emph{engineered thermal} distribution. Importantly, the process $\mathcal{L}_{\rm cool}$ strictly conserves the total particle number. Hence, in this idealized limit, the total average photon number in the steady state,
\begin{equation}\label{eq:Nfixed}
N_{\rm ph}(t\rightarrow \infty)=N_0 ,
\end{equation}
is determined by the initial photon number $N_0$. This condition also fixes the effective chemical potential $\mu_\mathrm{eff}$ in  Eq.~$\eqref{eq:efTh}$ through the constraint $\sum_k n_k\lvert_{\gamma=0}=N_0$.

For $\gamma,g\neq0$ the mode occupation numbers are determined by the interplay between the two competing thermalization mechanisms. In the following we characterize this competition by the dimensionless parameter $G=\gamma\Gamma/g^2$, which corresponds to the ratio of rates for the naturally occurring and the engineered dissipation processes for the fundamental mode, $k=1$.
The presence of both processes means that the photon-number conservation in Eq.~\eqref{eq:Nfixed} does not hold anymore. Remarkably, however, we find that this combined process 
%still gives rise to an effective conservation law for the 
\rev{can still be understood in terms of a constraint on the }\textit{energy} of the steady state. To understand this last point, we notice that changes in the total particle number, ${\dot{N}_{\rm ph}=\sum_k \dot{n}_k=-\sum_k \gamma_k(n_k-n_k^{\text{th}}})$, occur only through linear losses described by Eq.~\eqref{eq:Lloss}. In the steady state, this net dissipation vanishes, which imposes a global constraint on the stationary distribution.
Specifically, for the considered Ohmic bath with $\gamma_k\sim \omega_k \sim k$, the condition $\dot{N}_{\rm ph}=0$ becomes
\begin{equation}\label{eq:Efixed}
E_{\rm ph}(t\rightarrow \infty)=E^\text{th}=\sum_k \hbar\omega_k n_k^{\text{th}} ,
\end{equation}
independently of the values of both $g$ and \rev{$\gamma>0$}. This difference between the conserved quantities in Eq.~\eqref{eq:Nfixed} and Eq.~\eqref{eq:Efixed} shows that even an infinitesimal coupling to the environment can drastically change the nature of the distribution. Despite the engineered cooling process, the total energy of the 1D photon gas in steady state is the same as the energy of the equilibrium configuration, and is only determined by the support temperature $T$. Thus, cooling in this setup must be associated with an increase of the total photon number, which reduces the energy per particle, $E_{\rm ph}/N_{\rm ph}$.  

\rev{This increase of the total photon number can be intuitively understood as follows. For an  Ohmic bath, the probability for a photon to enter or exit the waveguide is proportional to its energy. In equilibrium, the steady state populations $n^{\rm th}_k$ arise from a balance of slow gain and slow losses in low-$k$ modes, and fast gain and losses in high-$k$ modes. When the engineered cooling mechanism is switched on, it shuttles high-energy photons into low-energy modes, where their decay rate is much lower, and where they can  thus become stuck for a long time. This effectively increases the average lifetime of photons for a fixed thermal injection rate, thus increasing the total population.}

\begin{figure}
    \centering
    \includegraphics[width=\linewidth]{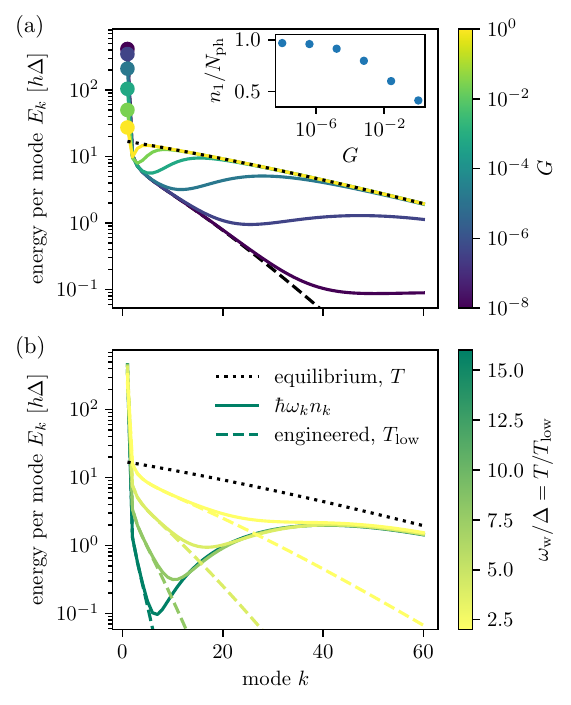}  
    \caption{Typical steady-state distributions of the average mode energies $E_k$  for a temperature of $T=T_{\rm w} =50\,$mK and a mode spacing of $\Delta/(2\pi)=60$ MHz, which corresponds to $k_BT/(\hbar \Delta)\approx 20$. (a) Energy per mode for different values of the dissipation parameter $G=\gamma \Gamma/g^2$ and $\Delta/\omega_\mathrm{w}=1/3$. The dotted and the dashed lines indicate the behavior the Bose-Einstein distributions given in Eq.~\eqref{eq:BE_eq} and Eq.~\eqref{eq:efTh} for temperatures $T$ and $T_\text{low}$, respectively. The inset shows the occupation of the lowest mode as a function of $G$. As this ratio is reduced, almost all photons end up in the ground mode. (b) Plot of the steady-state energy distribution for varying  $\omega_{\rm w}/\Delta=T/T_{\rm low}$ and for a fixed value of $G=10^{-6}$. }
    \label{fig:dists}
\end{figure}

In Fig.~\ref{fig:dists}(a), we plot the steady-state distribution of the energy per mode, $E_k=\hbar \omega_k n_k$,  for \rev{six different values of $G$ uniformly distributed on a logarithmic scale between $G=1$ and $G=10^{-8}$} and for $T_{\rm w}=T$. For a thermal Planck distribution,  $E_k$ appears as an almost straight line in this logarithmic plot. 
\rev{The dashed lines indicate the steady-state distribution in the limit $G\rightarrow0$, which coincides with the engineered thermal distribution given in Eq.~\eqref{eq:efTh} for an effective chemical potential $\mu_{\rm eff}$ which is determined by the condition in Eq.~\eqref{eq:Efixed}. Note that for any of the intermediate distributions, a chemical potential is not defined.}   
Generally, the lowest modes tend to thermalize at the engineered temperature $T_\text{low}$, while the higher modes thermalize at the environment temperature $T$. Due to the constraint on the total energy, $E_{\rm ph}=E^\text{th}$, the energy that is removed from the modes with $k\geq 2$ is redistributed to the fundamental mode and $n_1>n_1^\mathrm{th}$. The amount of redistributed energy and, consequently, both the fraction $n_1/N_{\rm ph}$ \rev{and the number of modes thermalized at $T_{\text{low}}$, are} controlled by the dissipation ratio $G$. By lowering $G$ (either by reducing the coupling to the thermal environment or by increasing the engineered cooling rate), one can increase the occupation of the lowest mode until it saturates at a value of $(n_1/N_{\rm ph})|_{\gamma\rightarrow0}<1$, which is determined by the effective thermal distribution given in Eq.~\eqref{eq:efTh}. In all intermediate distributions, the detailed balance between neighboring modes is weakly broken, and even in the steady state an effective flow of photons from high-frequency modes to low-frequency modes is established. This highlights the non-equilibrium nature of these distributions.  

In Fig.~\ref{fig:dists}(b), we vary the engineered temperature \rev{$T/T_{\rm low}=2,4,8,16$} by changing the frequency of the waste mode \rev{for fixed $G=10^{-6}$}. When increasing $\omega_{\rm w}$ (i.e., decreasing $T_{\rm low}$), the redistribution of the energy is restricted to a smaller and smaller number of modes, which, however, thermalize at an ever-lower temperature. Hence, although $T_{\rm low}$ may in principle be arbitrarily low, the constraint on the total energy restricts the number of modes that can be cooled to this lower temperature. In Appendix~\ref{app:CooledModes} we present an estimate for the number of efficiently cooled modes, which shows that this number only depends weakly on the dissipation ratio $G$.

\subsection{Effective temperature}\label{subsec:TeffDerivation}

The observed stabilization of multiple modes at a temperature that is below that of the surrounding bath, combined with an energy transfer toward the lowest mode, clearly indicates that the engineered dissipation mechanism "cools" the photon gas. However, the steady state is a nonequilibrium distribution that cannot be easily described by a single temperature; it is therefore not obvious how the cooling performances of this setup can be quantified.
Here we adopt an operative approach that characterizes the non-equilibrium photon gas in terms of its ability to exchange energy with another ancillary system. In other words, we are interested in the effective temperature $T_{\rm eff}$ that another system would thermalize to when weakly coupled to the photon gas. 

To arrive at a meaningful definition for $T_{\rm eff}$, we make a Gedanken experiment and consider a (fictitious) system with an arbitrary level structure, which exchanges energy with our photonic gas. Specifically, we assume a weak, nonlinear coupling of the form 
\begin{equation}
    V=\eta\sum_{\omega,k,k'}a_k^\dagger a_{k'}\sigma^+_\omega+ {\rm H.c.},
\end{equation}
where the operator $\sigma^+_\omega$ describes transitions between states of the ancillary system with an energy difference of $\hbar\omega$.
Similar to the coupling between the photon gas and the waste mode, this process conserves the total photon number in the waveguide, but permits energy exchange. In the limit $\eta\rightarrow0$, we can model this energy transfer by treating the photon gas as a Markovian reservoir and derive a master equation for the reduced state of the ancillary system,
\begin{align}\label{eq:mbCool}
    \dot{\rho}_{\rm a}= \Gamma_{\rm a}{\sum_{\omega,k,k'}}^{\prime} \Big\{&n_k(n_{k'}+1)\mathcal{D}[\sigma_\omega^-]\\ 
    \notag
    &+(n_k+1)n_{k'}\mathcal{D}[\sigma_\omega^+]\Big\}\rho_{\rm a}.
\end{align}
Here $\sum^\prime$ denotes the sum over pairs of modes that conserve the energy, i.e., $\rev{\omega_{k'}-\omega_{k}}\approx\omega$ and $\Gamma_{\rm a}$ is an effective damping rate, which we assume to be frequency-independent and whose precise value is unimportant for the present argument.

To proceed, we compare Eq.~\eqref{eq:mbCool} with the standard form of a master equation for a system coupled to a thermal reservoir, $\dot{\rho}=\sum_\omega\lambda_\omega\mathcal{D}[\sigma_\omega^+]\rho+\mu_\omega\mathcal{D}[\sigma_\omega^-]\rho$, where the absorption and emission rates at each frequency are related through $\lambda_\omega/\mu_\omega=e^{-\beta\hbar\omega}$. Hence, for a given transition frequency $\omega_k=k\Delta$, we can define an effective temperature $T^{\omega_k}_\text{eff}$ as
\begin{equation}
    \frac{1}{T^{\omega_k}_\text{eff}}=\frac{k_B}{\hbar\omega_k} \log\left(\frac{\sum_q (n_{q+k}+1)n_{q}}{\sum_q (n_q+1)n_{q+k}}\right)
\end{equation}
\rev{where, importantly,  the sums over $q$ include all modes up to ${k+q=k_{\rm max}}$. Hence, this quantity is a global property of the distribution, which involves all modes in the waveguide.} We have thus achieved a characterization of our steady-state distribution through its ability to thermalize a system with a single characteristic frequency.

In general, we are interested in thermalizing a complex system with many energy levels and associated transitions over a characteristic spectral range $[\Delta, K\Delta]$. We can then define an effective temperature over this spectral range as
\begin{equation}\label{eq:Teff}
    T_{\rm eff}=\frac{1}{K} \sum_{k=1}^K T_{\rm eff}^{\rev{\omega_k}},
\end{equation}
which recovers the limit $T_{\rm eff}\rightarrow T_{\rm low}$ from Eq.~\eqref{eq:efTh} as $\gamma\rightarrow0$ and $T_{\rm eff}\rightarrow T$ in the opposite regime \rev{and is, again, a function of all mode occupations $n_{1,\dots, k_{\rm max}}$}. 
Therefore, this operative approach allows us to characterize the nonequilibrium distribution in terms of a single effective temperature, which can be interpreted as the average temperature any other systems would experience, when being weakly coupled to the photon gas.  This effective temperature still depends on the spectral range $K\approx E_{\rm a}/(\hbar \Delta)$, which is in turn determined by the characteristic energy scales up to which the ancilla system can efficiently absorb excitations. Therefore, we obtain a well-defined temperature for the nonequilibrium photon gas, once the context of the experiment---in the current example the relevant energy interval $[0,E_{\rm a}]$---is specified.    

\subsection{Cooling performance}

\begin{figure}
    \centering
    \includegraphics[width=\linewidth]{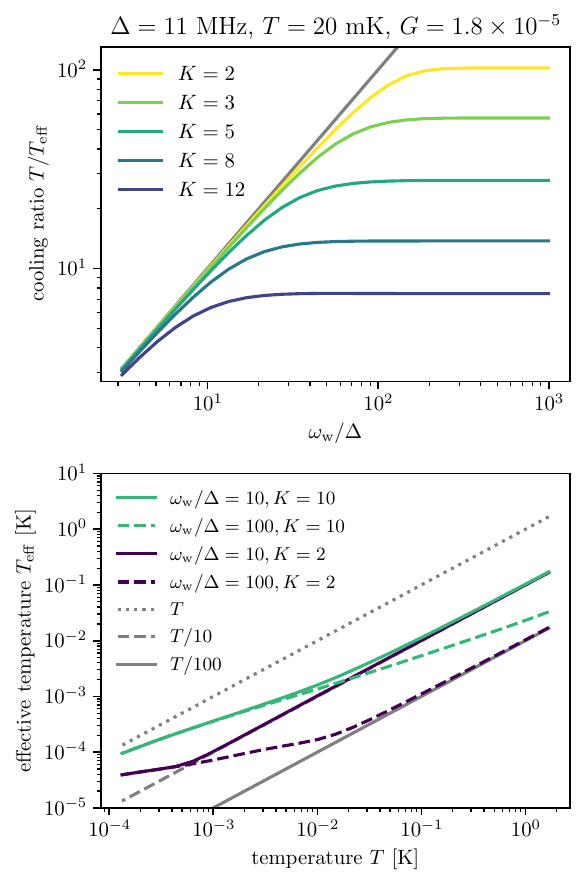}
    \put(-243,350){(a)} 
    \put(-243,170){(b)}
    \caption{Performance of the engineered cooling scheme for a representative set of experimental parameters and quantified in terms of the effective temperature $T_{\rm eff}$ introduced in Eq.~\eqref{eq:Teff}. (a) Plot of the cooling factor $T/T_{\rm eff}$ as a function of $\omega_{\rm w}/\Delta$ and for different spectral ranges $K$. \rev{The grey straight line corresponds to $T=T_{\text{low}}$.} (b) Plot of the effective temperature as a function of the support temperature and for different values of $\omega_{\rm w}/\Delta$ and $K$.  }
    \label{fig:cool}
\end{figure}

Based on the definition of the effective temperature $T_{\rm eff}$ motivated above, we can quantify the performance of the cooling scheme in terms of the ratio $T/T_{\rm eff}$. In Fig.~\ref{fig:cool}(a), we plot the achievable cooling ratio as a function of the waste frequency $\omega_{\rm w}/\Delta$ and spectral range $K$, assuming $T_{\rm w}=T$. \rev{The grey solid line indicates $T_{\rm eff}=T_{\rm low}$.} We observe two distinct regimes. For modest $\omega_{\rm w}/\Delta > 1$ and small spectral range, all of the modes thermalize close to the temperature $T_\text{low}$ defined by the engineered dissipation mechanism, see also Fig.~\ref{fig:dists}. The effective global temperature is therefore also given by $T_\text{eff}\approx T_\text{low}=T\Delta /\omega_{\rm w}$, and the cooling ratio $T/T_\text{eff}$ increases linearly with $\omega_{\rm w}/\Delta$. However, for very large values of the waste mode frequency, only the first few modes thermalize at $T_\text{low}$, whereas modes with higher $k$ settle at increasingly higher temperatures, such that the effective temperature saturates at $T_\text{eff}\approx T$ when considering a sufficiently large spectral range, i.e. $K\gg1$. Note that in this regime, i.e., when $n_{\rm w}\ll 1$, the populations $n_k$ start to deviate from the detailed-balanced one in Eq.~\eqref{eq:efTh} already for $k\geq 2$ (see example distributions in Appendix~\ref{app:dists}). This can be understood from the onset of a zig-zag instability in directional bosonic transport~\cite{Garbe2024}. \rev{For $T_{\rm low}=1$ mK, which corresponds to $\omega_{\rm w}/\Delta=20$, roughly five modes are fully thermalized to $T_{\rm low}$.}

In Fig.~\ref{fig:cool}(b) we plot the absolute value of $T_{\rm eff}$ as a function of the support temperature $T$ and for different ratios $\omega_{\rm w}/\Delta$. We see that, over a large temperature range, cooling factors of $10-100$ can be reached, depending on the spectral range of interest. Therefore, for the considered parameters and starting from a dilution refrigerator temperature of about $T=20$ mK, the few lowest modes of the photon gas can be cooled to effective temperatures of $0.1-1$ mK, while retaining a finite number of about $N_{\rm ph}\approx750$ photons in the gas. We can compare this with an equilibrium photon gas at the same temperature, for which we would have $N_{\rm ph}\approx150$. 

\rev{Note that generally the photon temperature can be higher than the base temperature of the fridge \cite{Yeh2017,Wang2019} if the photon field is heated by control lines. However, the condensate studied here can be isolated during its dissipative stabilization which allows for the assumed temperature of $T=20\;$mK}.

\section{Nonequilibrium condensation in a 1D transmission line}
One of the most salient features of the distributions shown in Fig.~\ref{fig:dists} is the accumulation of a large number of photons in the lowest mode, indicating condensation. This occurs despite the fact that, for a 1D Bose gas with a dispersion relation $\omega_k\propto |k|$, no condensation is expected in the thermodynamic limit when the system is in equilibrium~\cite{AguileraNavarro1999}.
In the following, we explore in more detail the origin and further properties of this phenomenon in the current, non-equilibrium scenario. To do so we focus on the idealized limit $G\rightarrow 0$, and assume that to a good approximation all relevant modes thermalize at the engineered temperature $T_{\rm low}$, i.e., the steady-state distribution is fully described by Eq.~\eqref{eq:efTh} [corresponding to the dashed lines in Fig.~\ref{fig:dists}]. 
In this section, we will also keep the waste mode temperature $T_{\rm w}$ as an independent parameter. 

\subsection{Condensate fraction}
As explained above, in the considered setup with an Ohmic coupling to the environment, the stationary photon distribution is constraint by the condition $E_{\rm ph}=E^\mathrm{th}$.
In the thermodynamic limit, $\Delta\rightarrow 0$ (or $\ell\rightarrow\infty$), we can approximate the sums in that condition by Bose integrals and we obtain
\begin{align}
\int_1^\infty\frac{k\,dk}{e^{\beta \hbar\omega_k}-1}= n_1+ \int_2^\infty\frac{k\,dk}{e^{\beta_{\rm low} (\hbar\omega_k-\mu_{\rm eff})}-1}.
\label{eq:fundmodepop}
\end{align}
Crucially, this energy constraint implies that the total photon number must increase when $T_{\rm low}<T$ and that these photons are added to the lowest mode, which is thus treated separately. This property distinguishes the effectively cooled photon gas from an equilibrium gas of massive bosons, which is cooled under constant photon number, and it is the reason why we can achieve a condensation phase transition. 

For any $T_{\rm low}<T$ the effective chemical potential $\mu_{\rm eff}$ in the right-hand side of Eq.~\eqref{eq:fundmodepop} lies between $0$ and $\hbar \omega_1$ and only gives rise to a correction of order \rev{$\mathcal{O}(\beta\hbar\Delta)$}. By neglecting this correction and solving Eq.~\eqref{eq:fundmodepop} in the thermodynamic limit, we see that the occupation number of the lowest photon mode,
\begin{equation}
n_1=\frac{\pi^2 }{6}\left(\frac{k_BT}{\hbar \Delta}\right)^2\left[1-\left(\frac{T_{\rm low}}{T}\right)^2\right],
\label{eq:groundpop}
\end{equation} 
indeed assumes a macroscopic value. From the occupation of the condensate mode, we can then also derive the effective chemical potential using Eq.~\eqref{eq:efTh}, and we obtain

\begin{equation}
        \mu_{\rm eff}\approx\hbar\omega_1-\frac{6}{\pi^2 }\left(\frac{\hbar\Delta}{k_B T}\right)^2
    \frac{k_B T_{\rm low}}{1-\left(T_{\rm low}/T\right)^2}.
\end{equation}
Hence, for any cooling ratio $T_{\rm low}/T<1$, the chemical potential approaches the ground-state energy, ${(\mu_{\rm eff}-\hbar\omega_1)/(\hbar \Delta)\rightarrow0}$ when $\Delta\rightarrow0$, and the occupation $n_1\rightarrow\infty$ diverges. To assess that we have condensation, this divergence must be compared to the number of photons in the excited modes, ${N_{\rm ex}=\sum_{k\ge2}n_k}$, which can be approximately evaluated by the integral 
\begin{align}\label{eq:Nph}
N_\text{ex}&\simeq \int_2^\infty\frac{dk}{e^{\beta_{\rm low}(\hbar \omega_k-\mu_{\rm eff})}-1}\notag\\
&=\left(\frac{k_B T_{\rm low}}{\hbar \Delta} \right)\ln\left(\frac{k_B T_{\rm low}}{\hbar \Delta} \right)+\mathcal{O}\left(\frac{k_B T_{\rm low}}{\hbar \Delta} \right).
\end{align}
Importantly, although the number of excited photons $N_{\rm ex}$ diverges when $\Delta\rightarrow 0$, as expected also for an equilibrium  gas of photons in 1D, this divergence is slower than the increase in $n_1$ evaluated in  Eq.~\eqref{eq:groundpop}. Therefore, in the thermodynamic limit the condensate fraction $n_1/N_\mathrm{ph}=n_1/(n_1+N_{\rm ex})$ reaches $1$ for any value $T_{\rm low}/T<1$. \rev{Note that a similar result will also follow for $\gamma_k\sim k^\alpha$, as long as $\alpha>0$.} 

In the opposite case when $T_{\rm low}>T$, the energy constraint given in Eq.~\eqref{eq:fundmodepop} can only be satisfied for a negative effective chemical potential, which approaches a value of 
\begin{equation}
    \beta_{\rm low}\mu_{\rm eff}\simeq\ln\left(\frac{\pi^2}{6}\frac{T^2}{T_{\rm low}^2}\right)
\end{equation}
in the limit $T_{\rm low}/T\gg1$. For any negative $\mu_{\rm eff}$, the condensation fraction $n_1/N_{\rm ph}$ vanishes in the thermodynamic limit.

\begin{figure}
	\centering
	\includegraphics[width=\linewidth]{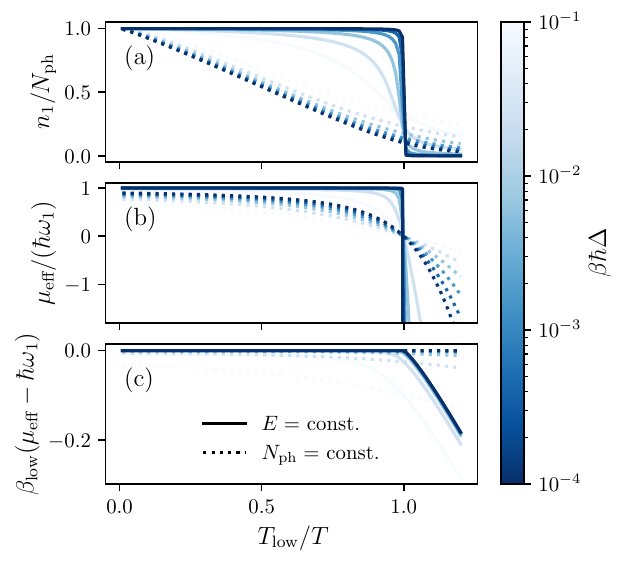}
	\caption{Nonequilibrium condensation of microwave photons. The solid lines show the exact numerical results for the condensation fraction $n_1/N_{\rm ph}$ and the effective chemical potential $\mu_{\rm eff}$ for an increasing length $\ell$ of the transmission line (decreasing mode spacing $\Delta$) and as a function of $T_\text{low}/T=T_{\rm w}\Delta/(T \omega_\text{w})$.
    The dotted lines show the corresponding quantities for a 1D gas of  bosons in equilibrium at temperature $T_{\rm low}$ and with a fixed particle number $N_{\rm ph}=N^{\rm th}$, which does not exhibit a sharp transition.  In the thermodynamic limit $\ell\propto(\beta\hbar\Delta)^{-1}\rightarrow\infty$ the condensation transition of the photon gas leads to a step function in the condensation fraction plotted in panel (a), and a sudden increase of the chemical potential from a large negative value to $\mu_{\rm eff}=\omega_1$, which is shown in panel (b) and in panel (c). 
    For all plots, the idealized limit $G=0^+$ has been assumed, in which case the nonequilibrium occupation numbers are determined by Eq.~\eqref{eq:efTh}.   
    }
	\label{fig:condens}
\end{figure}

In Fig.~\ref{fig:condens} we illustrate this behavior for the condensation fraction and the effective chemical potential in a finite system. Darker lines correspond to larger system sizes $\ell\propto(\beta\hbar\Delta)^{-1}$ and approach the thermodynamic limit. \rev{The six displayed values of $\beta\hbar\Delta$ are uniformly distributed between $10^{-4}$ and $10^{-1}$ on a logarithmic scale.} For comparison, dotted lines show the behavior of a 1D equilibrium gas of massive bosons with a constant total particle number of $N_{\rm ph}=\sum_kn^{\rm th}_k$, which does not exhibit a sharp transition. The comparison of the two scenarios illustrates how the sudden rise of the chemical potential from $\mu_{\rm eff}<0$ to $\mu_{\rm eff}\simeq \omega_1$ around $T_{\rm low}\simeq T$ causes the photon number to rapidly increase and therefore induce a macroscopic occupation $n_1$ even though the Bose-Einstein distribution has the same temperature as the one describing massive bosons.

To tune the system across the condensation transition at $T_{\rm low}/T=(T_{\rm w}/T)(\Delta/\omega_{\rm w})=1$, one can vary the ratio $\Delta/\omega_{\rm w}$ and use an external driving field to compensate for the frequency mismatch (see Sec.~\ref{sec::impl}). Alternatively, one can vary the ratio $T_{\rm w}/T$ by coupling the transmission line or the waste mode to different parts of the circuit held at different local temperatures. For example, for $\Delta/(2\pi) =30$ MHz and $T=250$ mK, we obtain $\beta\hbar \Delta\approx 0.006$, for which according to Fig.~\ref{fig:condens} a rather sharp transition is expected. For $\omega_{\rm w}/(2\pi)\approx 250$ MHz, thermal radiation at a temperature of $T_{\rm w}\approx 2$ K, which is injected, for example, from a high-temperature stage of the dilution refrigerator, can be used to tune the system across the transition. Note that for finite values of $G$, not all of the modes thermalize to $T_{\rm low}$ and the transition will be less pronounced. However, even in this case the increase of the condensate fraction $n_1/N_{\rm ph}$ near $T\approx T_{\rm low}$ is still clearly distinct from that of a photon gas in equilibrium.

\begin{figure}
	\centering
	\includegraphics[width=\linewidth]{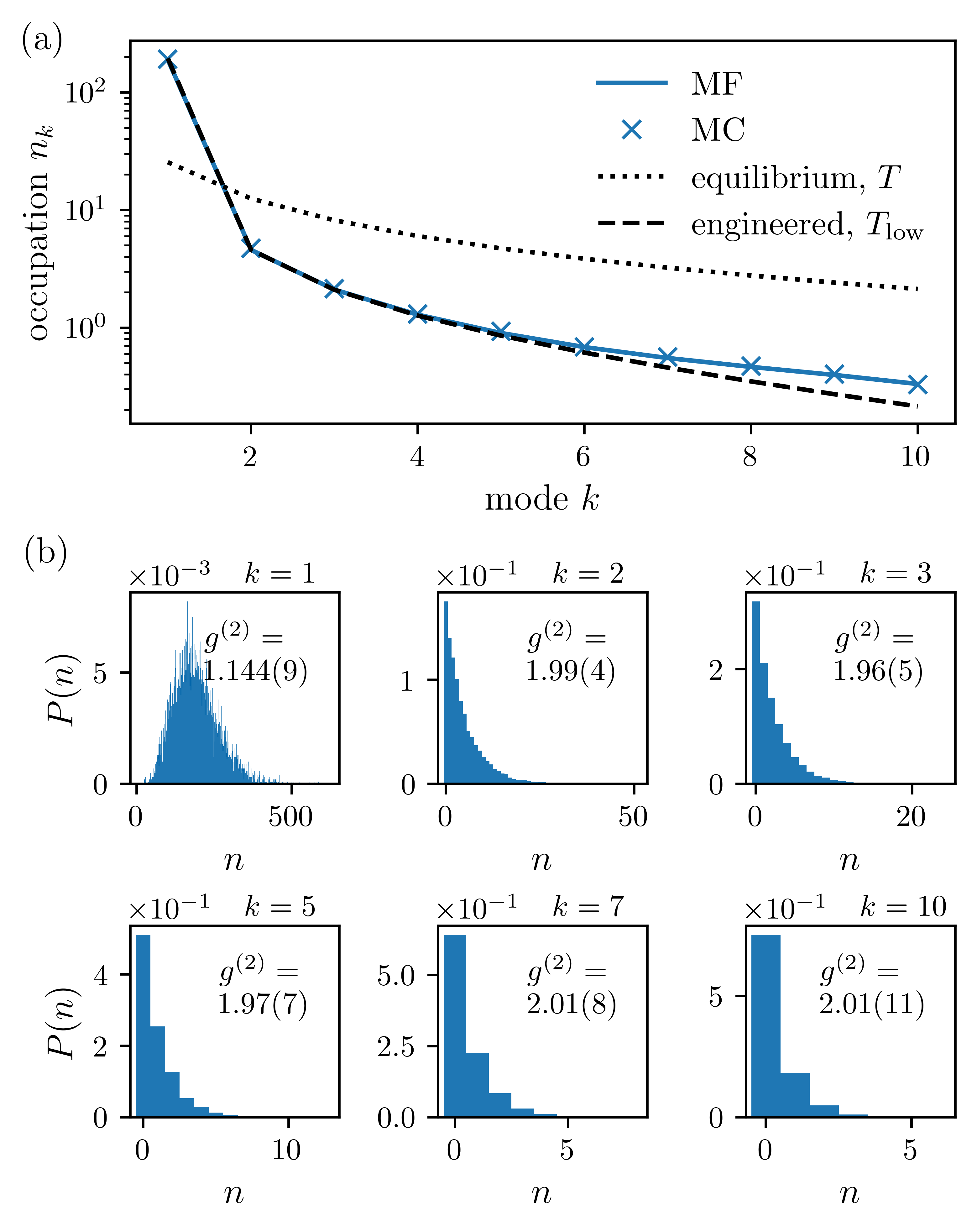}
    \caption{Photon-number distributions. (a) The average steady-state occupation numbers $n_k$ obtained from Monte Carlo simulations (MC) are compared with the corresponding mean-field (MF) results and with the equilibrium distributions given in Eq.~\eqref{eq:BE_eq} and Eq.~\eqref{eq:efTh}. 
    (b) Plots of the steady-state distributions $P(n)$ of photon number states $|n\rangle_k$ for a selection of modes, together with the corresponding value of the second-order correlation function $g^{(2)}(0)$. These distributions have been obtained from Monte Carlo simulations. For both plots, the same parameters $T=100\;$mK, $\Delta/(2\pi)=80\;$MHz, $G=10^{-4}$, $\omega_\mathrm{w}/\Delta=5$ and a cutoff of $k_{\rm max}=10$ have been assumed. }
	\label{fig:twa}
\end{figure}

\subsection{Condensate fluctuations}

To characterize the nonequilibrium state of the photon gas beyond the average occupation numbers, we apply Monte Carlo sampling in order to address the number statistics of the individual modes. Since the master equation is diagonal in the Fock basis $\ket{\{n_1,n_2,\dots,n_{k_{\rm max}}\}}$, the exact dynamics of the probabilities $P(\{n_k\},t)=\bra{\{n_k\}}{\rho(t)}\ket{\{n_k\}}$ in this configuration space are accessible. This supplements the mean-field approximation, since occupation number fluctuations and correlations between different modes are captured. For details on the algorithm we refer to Appendix~\ref{app:mc}.

We apply Monte Carlo sampling to simulate a small system of ten modes with the parameters $\Delta/(2\pi)=80\,$MHz, $G=10^{-4}$, $\omega_\text{w}/\Delta=5$ and $T=100\,$mK, \rev{using $10^4$ trajectories.} 
We calculate the mean occupation number and the second-order correlation function, 
\begin{equation}
g_k^{(2)}(0)= \frac{\langle a_k^\dag a^\dag_k a_k a_k\rangle}{\langle a^\dag_k a_k\rangle^2},
\end{equation}
for each mode. \rev{Sampling errors are given in brackets after the $g^{(2)}$-values and convergence errors are ensured to be small in comparison.} We find that the occupation numbers agree with the mean-field predictions with very high accuracy, as shown in Fig.~\ref{fig:twa}(a). 
In Fig.~\ref{fig:twa}(b), we plot the corresponding photon-number distributions for a representative selection of modes, together with the corresponding values of $g_k^{(2)}(0)$. For modes with $k\ge2$, the photon-number distribution is well approximated by an exponential distribution, consistent with a thermal state. This is also confirmed by the correlations, which exhibit a value of  $g^{(2)}(0)\simeq 2$, up to sampling errors. In the lowest mode, by contrast, the distribution resembles more closely a Poisson distribution, with a significantly lower value of $g^{(2)}(0)\simeq 1$, close to the value associated with a coherent state. \rev{The observed residual deviation from unity is expected for a condensate with a finite number of particles.} These results show that the nonequilibrium distribution resembles that of a  Bose-Einstein-condensate also at the level of second-order correlations.   
Similar to the analysis performed in Ref.~\cite{Garbe2024}, additional simulations of the first-order coherence (not shown) also confirm that there is a $U(1)$-symmetry breaking in the condensate mode.

\section{Implementation}\label{sec::impl}
In the setup depicted in Fig.~\ref{fig:model}, the 1D photon gas is realized by a superconducting transmission line of length $\ell$, group velocity $v$ and impedance $Z$,  which supports a set of standing-wave modes with frequency spacing $\Delta= v \pi /\ell $. These modes decay with rates $\gamma_k=\omega_k/Q$ into a thermal environment at temperature $T$.  One end of the transmission line is coupled to an additional $LC$-resonator with frequency $\omega_\text{w}$ and impedance $Z_\text{w}$. This waste mode dissipates energy with a rate $\Gamma$ into a thermal environment held at a temperature $T_{\rm w}$. For most of our analysis we have assumed $T_{\rm w}=T$, but in general $T_{\rm w}$ can be treated as an independent, locally adjustable temperature. To realize a three-wave mixing interaction as assumed in Eq.~\eqref{eq:Hint}, the waste mode is coupled to the transmission line via a SNAIL-type loop of Josephson junctions \cite{Frattini2018}, which is penetrated by a time-dependent external flux $\phi_{\rm ex}(t)$. 

\subsection{Three-wave-mixing Hamiltonian}

The Hamiltonian of the closed system is given by 
\begin{equation}
H=H_\text{wg}+H_\text{w}+H_{\rm c},
%(\varphi_\text{wg}-\varphi_\text{w}),
\end{equation}
 where the individual terms represent the Hamiltonians of the waveguide, the waste mode and the coupler, respectively. The Hamiltonian of the waveguide,
 \begin{equation}
H_{\rm wg}=\sum_{k=1}^{k_{\rm max}}\hbar\omega_k a_k^\dagger a_k^{\phantom{\dagger}},
\end{equation}
 consists of a set of harmonic modes and
 can be  derived from the discrete transmission line model following the standard procedure~\cite{GarciaRipoll2022}. In terms of these modes, the quantized flux operator at a position $x$ along the transmission line can be written as  
 \begin{align}
\phi(x)= \sum_k \phi_k^{0}\cos\left(\frac{kx\pi}{\ell}\right)\left(a_k+a_k^\dagger\right),
%&q_k=-iq_k^{0}\sqrt{2}\cos(kx\pi/\ell)(e^{-i\omega_kt}a_k-e^{i\omega_kt}a_k^\dagger).
\end{align}
where $\phi_k^{0}=\sqrt{\hbar Z/(\pi k)}$ are the zero-point fluctuations of each mode. The operator relevant for the engineered cooling process at the end of the waveguide is thus given by
\begin{equation}
\phi_\text{wg}\equiv\phi(x=0) =\sum_k\phi_k^{0}\left(a_k+a_k^\dagger\right).
\end{equation}
Similarly, the waste mode can be modeled as a single quantized harmonic mode with a Hamiltonian $H_{\rm w}=\hbar \omega_{\rm w} a_{\rm w}^\dag a_{\rm w}$ and a quantized flux operator $\phi_\mathrm{w}=\phi_\mathrm{w}^{0}(a_\mathrm{w}+a_\mathrm{w}^\dagger)$, where $\phi_\mathrm{w}^{0}=\sqrt{\hbar Z_\text{w}/2}$.

To obtain a tunable three-wave interaction between the photon gas and the waste mode, we consider a SNAIL-type coupler with three identical junctions with Josephson energy $E_J$ in parallel with one junction with Josephson energy $\alpha E_J$. By neglecting, for simplicity, the capacitive energy of the junctions, we can write $H_{\rm c}\simeq V_{\rm c}(\varphi_\text{wg}-\varphi_\text{w})$, 
where $\varphi_\text{wg}=\phi_\text{wg}/\phi_0$ and $\varphi_\text{w} =\phi_\text{w}/\phi_0$, $\phi_0=\hbar/(2e)$ is the reduced flux quantum and \rev{$e$ is the elementary charge}. In the presence of the external flux, $\varphi_{\rm ex}= \phi_{\rm ex}/\phi_0$, the resulting expression for the coupler potential reads
\begin{equation}\label{eq:expansion}
    \frac{V_{\rm c}(\varphi)}{E_J}=-\alpha\cos\left(\frac{\varphi_{\rm ex}}2+\varphi\right)-3\cos\left[\frac13\left({\frac{\varphi_{\rm ex}}2-\varphi}\right)\right].
\end{equation}
We remark that in order to express the potential in this symmetric form, we have chosen a specific time-dependent gauge, which gives rise to a charge drive after the Legendre transformation, if the junction capacitances $C_{Ji}$ do not satisfy a specific relation \cite{You2019}. The drive is, however, off-resonant and can be cancelled by applying additional driving fields. For alternative derivations of similar coupling schemes based on a different choice of gauge, see, for example, Refs.~\cite{Eriksson2024,Haider2024}.

By setting $\alpha=1/3$ and \mbox{$\varphi_\text{ex}=\pi+\varphi_t$}, where $\varphi_t$ is the time-dependent part of the flux, we achieve a potential that behaves as \mbox{$(\varphi-\pi/2)^4/81$} in leading order around its static minimum \rev{for $\varphi_t=0$. In the flux pumped case, the minimum is determined by the condition $V_c'(\varphi_m)=0$, which gives ${\varphi_m=\pi+\varphi_t/2+\mathcal{O}(\varphi_t^3)}$. We assume that the flux pump is small enough to expand the potential around that approximate minimum and find}
\rev{\begin{align}\label{eq:potential}\notag
V_{\rm c}(\varphi)=
&-3E_J\left(\cos\frac{\varphi_t}{6}\cos\frac{\varphi}{3}+\sin\frac{\varphi_t}{6}\sin\frac{\varphi}{3}\right)\\\notag
&+\frac13E_J\left(\cos\frac{\varphi_t}{2}\cos\varphi-\sin\frac{\varphi_t}{2}\sin\varphi\right)
\\
\;=&\;\mathrm{const.}+\frac{1}{54}E_J\varphi_t\varphi^3+\frac{1}{81}E_J\varphi^4\notag\\
&+\mathcal{O}(\varphi_t^2\varphi^2,\varphi_t^3\varphi,\varphi_t\varphi^5,\varphi^6,\dots),
\end{align}
}where for the truncation of the potential at cubic order we require $\varphi,\varphi_t\ll1$. %A similar mechanism has been implemented for single photon detection \cite{Nojiri2024}.

To obtain the desired interaction, we choose a periodic flux drive $\varphi_t=\epsilon\cos(\omega_\text{d}t)$, whose frequency $\omega_{\rm d}=\omega_\text{w}-\Delta$ is chosen to resonantly enhance interaction terms of the form $a_\mathrm{w}^\dagger a_k^\dagger a_{k+1}+{\rm H.c.}$ Simultaneously, we want to avoid any other resonances. For example, the sum of drive and waste frequencies should not couple any combination of widely separated levels $\omega_\text{w}+\omega_\text{d}\neq m\Delta$ for any integer $m$. Otherwise, there can be processes where one waveguide photon of any mode is absorbed by the waste mode via two-photon absorption. This process can be avoided by requiring $2\omega_\text{w}\pm\omega_\text{d}\neq m\Delta$, which can be satisfied, for example, by the choice
\begin{align}
&\omega_\text{w}=\left(\frac54 +\frac n2\right)\Delta,&\omega_\text{d}=\left(\frac14 +\frac n2\right)\Delta
\end{align}
for any $n=0,1,\dots$ Then, for a given support temperature $T=T_{\rm w}$, we can set $n\approx2T/T_{\rm low}-5/2$ to realize the targeted engineered temperature $T_{\rm low}$ in discrete steps. 

The requirement for the off-resonant couplings to be suppressed is that the linewidth of the waste mode, $\Gamma$, is smaller than $\Delta/4$, which is the minimal frequency spacing. By contrast, the ratio $\omega_\text{w}/\Delta$ can be made arbitrarily high by an appropriate choice of $\omega_\text{w}$ and $\omega_\text{d}$. Under this resonance condition and within the rotating-wave approximation, the interaction reduces to
\rev{\begin{align}\label{Hint}
H_{\rm c}\simeq & \frac{\epsilon E_J}{54} \cos(\omega_\mathrm{d}t)\left[\varphi_\text{w}^{0}(a_\mathrm{w}+a_\mathrm{w}^\dagger)-\sum_k\varphi^{0}_k(a_k+a_k^\dagger)\right]^3\notag\\
\simeq &\sum_k \frac{\hbar g}{\sqrt{k(k+1)}} \left(a_\mathrm{w}^\dagger a_k^\dagger a_{k+1} + a_\mathrm{w} a_k a^\dag_{k+1} \right),
\end{align}}
where we have introduced the coupling strength \rev{${g=\epsilon E_J\varphi_\text{w}^{0}(\varphi^{0}_1)^2/(18\hbar)}$}. This is the desired three-wave mixing Hamiltonian introduced in Eq.~\eqref{eq:Hint} and the starting point for the derivation of the master equation in Eq.~\eqref{eq:me}.

\subsection{Derivation of the master equation}\label{subsec:MEderivation}

Starting from the three-mode interaction in Eq.~\eqref{Hint}, we change into an interaction picture with respect to $H_0=H_{\rm wg} + H_{\rm w}$ and obtain
\begin{equation}
H_c=\hbar g\sum_k\frac{a_k^\dagger a_{k+1}a_\mathrm{w}^\dagger}{\sqrt{k(k+1)}} e^{it(\delta_k-\delta_{k+1})}+{\rm H.c.}
\end{equation}
Here we have included the detunings $\delta_k = \omega_k- k\Delta$, to allow for small deviations of the mode frequencies from being exactly equally spaced. Below we will use this generalized setting to estimate the sensitivity of the cooling scheme to Kerr effects and other potential frequency shifts. 

In the limit $g\ll \Gamma$, the quickly decaying waste mode can be treated as an effective Markovian bath for the waveguide modes and its dynamics can be adiabatically eliminated to obtain a master equation for the reduced state of the 1D photon gas only. By neglecting small frequency corrections in the Hamiltonian part, this master equation can be written in a compact form as
\begin{equation}\label{eq:FullME}
\begin{split}
\dot \rho = \sum_{k,k'} \kappa_{k} &\left\{(n_{\rm w}+1)  e^{i(\Omega_k- \Omega_{k'})t} [c_{k'}\rho, c^\dag_k]\right.\\
&\left. +  n_{\rm w} e^{-i(\Omega_k- \Omega_{k'})t} [c^\dag_{k'}\rho, c_k]+{\rm H.c.} \right\},
\end{split}
\end{equation}
where $c_k=a_k^\dagger a_{k+1}$, $\Omega_k=\delta_{k+1}-\delta_{k}$ and the 
\begin{equation}
\kappa_{k}=\frac{g^2}{k(k+1)} \frac{\Gamma}{\Gamma^2+\Omega_k^2}
\end{equation}
denote the effective rates.

We see that in an ideal system with $\Omega_k=0$, we obtain a time-independent master equation with a collective jump operator of the form $C=\sum_k a_k^\dagger a_{k+1}/\sqrt{k(k+1)}$. It describes cooling of all the modes, but in a correlated manner, since the waste mode is coupled to all transition $k+1\rightarrow k$ simultaneously. In practice, such a situation is very unlikely and small anharmonicities in the mode frequencies, decoherence effects or non-linear corrections from higher-order terms in the coupler will result in $\Omega_k\neq \Omega_{k'}$ and effectively eliminate the terms with $k\neq k'$ in the sum in Eq.~\eqref{eq:FullME}. Therefore, under the assumption 
\begin{equation}
\kappa_k \ll |\Omega_{k}-\Omega_{k'}| \ll \Gamma
\end{equation}
the master equation in Eq.~\eqref{eq:FullME} reduces to the engineered cooling master equation in Eq.~\eqref{eq:Lint}. Once the frequency shifts $\delta_k$ become too large, i.e., $|\Omega_k|\gtrsim \Gamma$, the corresponding modes will be out of resonance and will no longer contribute to the cooling dynamics. In practice, this will result in a finite cutoff $k_{\rm max}$ due to deviations from an equidistant mode spacing. This condition will also constrain the total number of photons in the condensed mode due a residual Kerr effect, as discussed in more detail in Appendix~\ref{app:parameters}.

\subsection{Experimental parameters}

The multi-mode cooling master equation in Eq.~\eqref{eq:Lint} has been derived under various assumptions about the system parameters, which are required, for example, to justify the expansion of the coupler potential in Eq.~\eqref{eq:expansion} or the adiabatic elimination of the waste mode in Sec.~\ref{subsec:MEderivation}. In Appendix \ref{app:parameters}, we provide a more detailed discussion of all those constraints and identify a regime of experimentally feasible parameters for which the derivation of our effective cooling master equation is valid. A typical set of parameters that satisfies  those constraints and leads to effective parameters similar to the ones used in our analysis above is summarized in Table~\ref{TabPara}.

In Appendix~\ref{app:parameters} we also propose a generalized coupler composed of a short array of $N_{\rm c}$ identical SNAIL elements. This approach can be used systematically reduce the Kerr-induced frequency shift of the fundamental mode, which arises from the subleading fourth-order term in the expansion in Eq.~\eqref{eq:expansion} and limits the total number of condensed photons. Note that such extensions and the parameters assumed in Table~\ref{TabPara} aim at the cooling of rather large, multi-mode systems. For proof-of-concept experiments with only a few modes, the thermal energy influx from all the higher modes is significantly reduced, and many constraints on the experimental parameters can be relaxed.

\begin{table}
\caption{Proposed experimental parameters for the refrigeration of a gas of microwave photons \rev{to temperatures of $T_{\rm low}\approx1$ mK-$20$ $\mu$K }.} 
\begin{ruledtabular}
\centering
\begin{tabular}{lcc}\label{TabPara}
Support temperature &$T$ & $20\;$mK \\
Mode spacing &$\Delta/(2\pi)$  & $11$ MHz\\
Quality factor &$Q$ & $5\times10^{6}$ \\
Waste mode frequency & $\omega_\mathrm{w}/(2\pi)$ & $200\;$MHz-$1$\;GHz \\
Waste mode decay rate& $\Gamma/(2\pi)$ & $1.8\;$MHz \\
Waste mode impedance &$Z_\mathrm{w}$ & $6\;\mathrm{k}\Omega$  \\
Waveguide impedance &$Z$ & $15\;\Omega$ \\
Josephson energy & $E_J/h$ & $30\;$GHz\\
Number of SNAILs & $N_{\rm c}$ & $3$ \\ 
Flux drive strength & $\epsilon$ & $0.4$\\
\rev{Cooling rate} & $g^2/\Gamma/(2\pi)$ & $120\;$kHz\\ 
Resulting dissipation ratio & $G$ & $1.8\times 10^{-5}$ 
\end{tabular}

\end{ruledtabular}
\end{table}

\section{Conclusion and Outlook}

In summary, we have proposed and analyzed a simple circuit QED setup to cool a 1D gas of microwave photons to sub-millikelvin temperatures and to prepare a photon condensate in a one-dimensional waveguide. We have shown that the steady state of this photon gas depends on the interplay between the engineered cooling processes and bare thermalization at the support temperature and that those two effects cannot be treated separately. In particular, we have identified an unexpected conservation of the total energy of the photon gas in this system, which replaces the conservation of the total number of bosons in gases of massive particles and induces a sharp condensation transition even in 1D.

In our analysis, we have focused on the case of a noninteracting photon gas as an illustrative example for the cooling a photonic many-body system in the microwave regime below the base temperature of a typical dilution refrigerator. \rev{While this system already shows interesting properties by itself, we envision that such an extremely cold gas of photons could be used in future applications as a universal coolant to prepare also more complicated many-body systems in such a low-temperature state.} Such applications are already anticipated in the operative definition of the effective temperature $T_{\rm eff}$ in Sec.~\ref{subsec:TeffDerivation}, which is defined as the effective temperature that another system  would thermalize at, when weakly coupled to the cold photon gas.

\begin{figure}
    \centering
    \includegraphics[width=\linewidth]{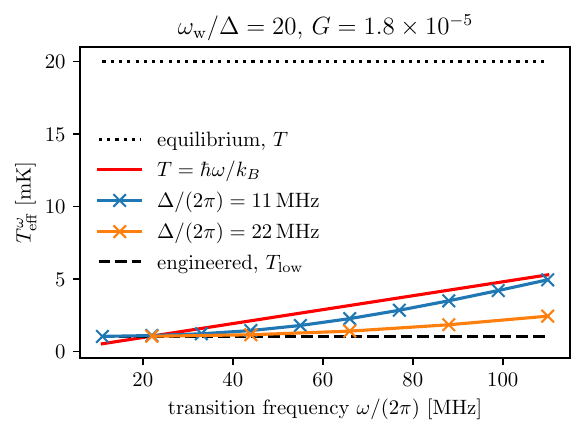}
    \caption{Many-body cooling using a pre-cooled microwave condensate. A transition with frequency $\omega$ of a generic many-body system is coupled to all resonant transitions in the 1D photon gas, which mimics a reservoir with a frequency-dependent effective temperature $T^\omega_{\rm eff}$.  The crosses show the values of $T^\omega_{\rm eff}$ obtained for a typical set of waveguide parameters and for two different mode spacings. \rev{The dashed line shows the target temperature $T_{\rm low}=1$ mK.} For those parameters, another many-body system with transitions up to $100$ MHz can be cooled to temperatures well below $5\;$mK. 
    }
    \label{fig:Teff2}
\end{figure}

As an explicit example, let us consider the implementation of a photonic Bose-Hubbard-like model~\cite{Ma2019}, with a typical on-site interaction energy of $U/h\approx 100$ MHz. This energy translates into a temperature of about $5$ mK, well below the usually achievable temperatures in a dilution refrigerator. In Fig.~\ref{fig:Teff2} we plot the (frequency resolved) effective temperature $T^\omega_{\rm eff}$ of the cooled photon gas for a realistic set of parameters and for transition frequencies $\omega$ up to $U/\hbar$. We see that within this frequency range, there is a range of possible transitions for which the effective temperature lies in the range of $1\,{\rm mK} < T^\omega_{\rm eff}< 5\,{\rm mK}$ and we obtain $T_{\rm eff}<2.5$ mK for the two chosen mode spacings. Therefore, by introducing a weak coupling between the two systems, the pre-cooled noninteracting photon gas can cool the interacting photons\rev{---at least in principle---}into a Mott-insulator state. \rev{We emphasize that this argument is solely based on ergodicity and the experimental feasibility of such sympathetic cooling schemes must still be verified through more involved many-body simulations of explicit circuit designs, which go beyond the scope of the current study. Nevertheless, this basic estimate already illustrates the prospects for using simple, pre-cooled photonic fluids to cool more complex photonic many-body system into a low-temperature equilibrium state with a fixed density.}

\begin{acknowledgments}
We thank Michael Haider, Adrian Parra-Rodriguez, Xin H.~H.~Zhang and Jacquelin Luneau for many helpful discussions and  acknowledge support by the Deutsche Forschungsgemeinschaft (German Research Foundation)–522216022. This research is part of the Munich Quantum Valley, which is supported by the Bavarian state government with funds from the Hightech Agenda Bayern Plus.
\end{acknowledgments}

\section*{Data availability}
\rev{The simulation code and numerical data are available at \cite{schamriss_2026}.}

\appendix

\section{Number of cooled modes}\label{app:CooledModes}
In this appendix, we derive an approximate estimate for the number of modes that are efficiently cooled to a mode occupation number $n_k\ll n_k^{\rm th}$ that is significantly below the equilibrium value. From the plots of the stationary energy distributions shown in Fig.~\ref{fig:dists}, we see that the crossover between the modes that thermalize at a temperature $T_{\rm low}$ and the higher modes that thermalize at the bath temperature $T$ occurs roughly at a value $k_c$, where the energy per mode has a local maximum, i.e., $\partial_kE_k=0$. \rev{To determine $k_c$, we can assume that the stationary mode occupation can be described by a Bose-Einstein distribution of the form
\begin{equation}
    n_k=\left(e^{\beta k \hbar\Omega_k}-1\right)^{-1}
\end{equation}
with an effective frequency $\Omega_k$ that slowly changes with $k$. For $k\ll k_c$, we have $\Omega_k\sim \omega_{\rm w}$, and for $k\gg k_c$, $\Omega_k\sim \Delta$. Provided that this $k-$dependence is slow enough, the relation 
\begin{equation}
 (1+n_k)n_{k+1}=n_\Omega (n_k-n_{k+1})  
\end{equation}
with $n_\Omega=(e^{\beta\hbar\Omega_k}-1)^{-1}$ holds for any temperature. The steady-state condition in Eq.~\eqref{eq:mf} then translates into the condition
\begin{equation}
    \kappa_k (n_{\rm w}-n_\Omega)(n_{k+1}+n_{k-1}-2n_k)=\gamma_k (n_k-n^{\rm th}_k),
    \label{eq:Sstate_condition}
\end{equation}
which we now evaluate in the vicinity of $k_c$. In this regime, we are still close to the engineered temperature, i.e.,  $\omega_{\rm w}-\Omega_{k_c}=\delta\omega\ll \omega_{\rm w}$. Expanding the term $n_{\rm w}-n_{\Omega}$ in Eq.~\eqref{eq:Sstate_condition} as a function of $\delta\omega$, we obtain $n_{\rm w}-n_{\Omega}\simeq -\beta\hbar\delta\omega/\{2[\cosh(\beta\hbar\omega_{\rm w})-1]\}$. Furthermore, the data shown in Fig.~\ref{fig:dists} indicates that around the point where the distribution starts deviating from the engineered thermal distribution at $T_{\rm low}$, the mode population is already much smaller than one, and also much below the equilibrium value. In other words, we can safely assume $n_{k_c}\ll n^{\rm th}_{k_c}$,  and approximate both populations around $k_c$ by thermal distributions $n_{k_c}\sim e^{-\beta \hbar\Omega_{k_c} k_c}\sim e^{-\beta\hbar\omega_{\rm w} k _c}$, and $n^{\rm th}_{k_c}\sim e^{-\beta\hbar\Delta k _c}$. This allows us to rewrite the first term in Eq.~\eqref{eq:Sstate_condition} as
\begin{equation}
(n_{k+1}+n_{k-1}-2n_k)\simeq \left[2\cosh(\beta
\hbar\omega_{\rm w})-1\right]n_k.
\end{equation}
Together with all other approximations for $k\sim k_c$, we then obtain 
%for $k\sim k_c$:%$\delta\omega_c=\Omega_{k_c}-\omega_w$:
\begin{equation}   \label{eq:condSstate} 
\hbar \delta\omega\simeq\frac{\gamma_k}{\beta\kappa_k}\frac{n^{\rm th}_{k}}{n_{k}}\sim \frac{Gk^3}{\beta}e^{\beta k\hbar(\omega_{\rm w}-\Delta)}
\end{equation}
To determine the inflection point of the distribution, $k_c$, we use the condition $\partial_k(kn_k)=0$, which gives
\begin{align}    \label{eq:condderiv}
\Omega(k_c)+k_c\partial_k\Omega\simeq 0\rightarrow \partial_{k} \delta\omega=\frac{\omega_{\rm w}}{k_c}.
\end{align}
Combining Eq.~\eqref{eq:condderiv} and Eq.~\eqref{eq:condSstate}, and using $\beta\hbar(\omega_{\rm w}-\Delta)k_c\gg1$, we finally arrive at an implicit equation for $k_c$, 
\begin{equation}
Gk_c^4=\frac{\omega_{\rm w}}{\omega_{\rm w}-\Delta}e^{-\beta\hbar(\omega_{\rm w}-\Delta)k_c}.
\end{equation}
}In the asymptotic limit 
${k_c\gg1}$, the solution of this implicit equation can be expressed as
\begin{equation}\label{eq:lambert}
k_c\approx\frac{4W\left(\frac14a^{\frac14}\beta\hbar(\omega_\text{w}-\Delta)\right)}{\beta\hbar(\omega_\text{w}-\Delta)},
\end{equation}
where $W(x)$ is the Lambert function and $a=e^{\beta\hbar\omega_\text{w}}/(1-\Delta/\omega_\text{w}) /G$. Since $W(x)$ diverges logarithmically for $x\rightarrow\infty$ and $a^{1/4}\propto G^{-1/4}$, the crossover mode number $k_c$ has only a weak dependence on the engineered cooling rate.

\section{Details of the steady-state distributions}\label{app:dists}

\begin{figure}
    \centering
    \includegraphics[width=\linewidth]{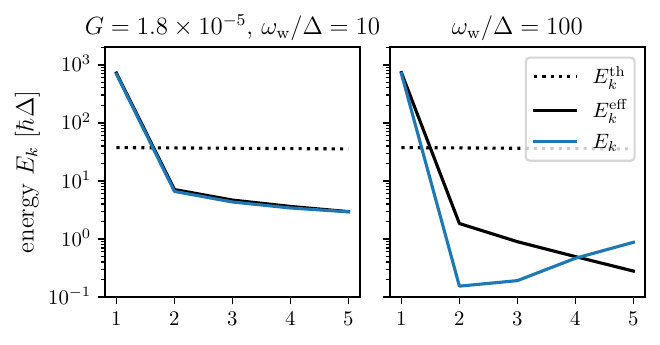}
    \includegraphics[width=\linewidth]{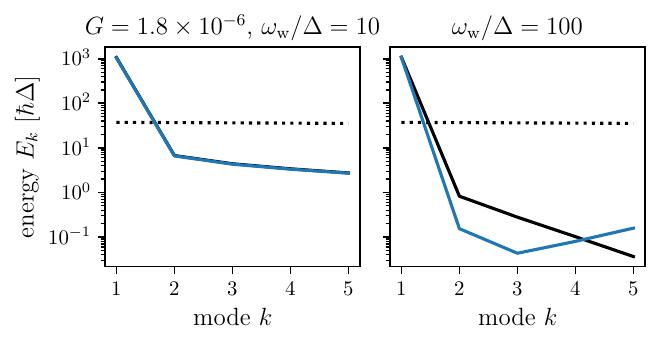}
    \caption{Examples of steady-state distributions related to Fig.~\ref{fig:cool}. The blue solid lines show the steady-state mode energies $E_k$, while the black solid line represents the corresponding values $E_k^{\rm eff}$ as obtained from a Bose-Einstein distribution with temperature $T_{\rm eff}$ for $K=5$ and a total photon number matching $N_{\rm ph}$. The dotted line indicates the thermal energy values $E_k^{\rm th}=\hbar\omega_kn_k^{\rm th}$ for reference. The upper and the lower panels differ in the assumed value for $G$. All other parameters are the same as in Fig.~\ref{fig:cool}. }
    \label{fig:cooling}
\end{figure}

In Fig.~\ref{fig:cooling} we provide additional plots for the stationary distribution of the mode energies, $E_k$, for some of the parameters assumed in  Fig.~\ref{fig:cool}. These plots illustrate the two regimes discussed in the main text. For $\omega_{\rm w}/\Delta=10$ the mode occupation numbers (blue) match that of a Bose-Einstein distribution (black) \rev{with an  effective temperature as given by Eq.~\eqref{eq:Teff} for $K=5$ and a chemical potential $\mu$ chosen according the condition $N_{\rm ph}=\sum_{k=1}^{k_{\rm max}}(e^{\hbar(\omega_k-\mu)/k_BT_{\rm eff}}-1)^{-1}$.} In contrast, for $\omega_{\rm w}/\Delta=100$ there are clear deviations already at small values of $k$. As $G$ is decreased, these deviations become less pronounced. We attribute the deviation from a thermal distribution for $n_{\rm w}\ll1$ to the partial formation of a staggered population profile, which is known from other bosonic systems with a directional flow~\cite{Liew2013,Garbe2024}.

\section{Monte-Carlo simulations}\label{app:mc}
In this section, we provide technical details about the Monte-Carlo simulations used to calculate number statistics. The algorithm runs $N$ stochastic trajectories starting from an arbitrary initial configuration $\{n_k\}_{t=0}$. In every step the time and the configuration of each trajectory is updated following the Gillespie algorithm \cite{Gillespie1977}, using two uniform random numbers $r_{1,2}\in (0,1]$. For a number of jump operators $L_j$ and corresponding rates $a_j$ the time is updated according to 
\begin{align}
    t_n&=t_{n-1}-\frac{\ln{r_1}}{\sum_j a_j}
\end{align}
and the configuration is updated using a jump operator $L_{i_n}$ which is chosen at random with probability $a_{i_n}/\sum_ja_j$ using $r_2$. The update is defined by
\begin{align}
    \ket{\{n_k\}_n}=\frac{L_{i_n}\ket{\{n_k\}_{n-1}}}{\|L_{i_n}\ket{\{n_k\}_{n-1}}\|}.
\end{align}
The configurations $\ket{\{n_k\}_n}$ obtained from the algorithm follow exactly the probability distribution $P(\{n_k\},t_n)$ as given by the diagonal density matrix evolved according to the master equation with the initial condition ${\rho_{t=0}=\ket{\{n_k\}_{t=0}}\bra{\{n_k\}_{t=0}}}$. Any higher statistical moments can be estimated with ensemble averages over Monte-Carlo trajectories. In particular for the second-order zero-time autocorrelation of the mode $k$ for samples $n_k^{(1,\dots,N)}$ one finds
\begin{equation}
    g^{(2)}_k(0)=\frac{\langle a_k^\dagger a_k^\dagger a_ka_k\rangle}{\langle a_k^\dagger a_k\rangle^2}\approx \frac{\frac1N\sum_n \left[\left( n^{(n)}_k\right)^2-n^{(n)}_k\right]}{\left(\frac1N\sum_n n_k^{(n)}\right)^2},
\end{equation}
which becomes exact in the limit $N\rightarrow\infty$.

\section{Experimental parameters}\label{app:parameters}
In the derivation of the three-mode interaction in Sec.~\ref{sec::impl}, we have neglected higher-order terms in the expansion of $V_{\rm c}$, assuming that $\varphi_{\rm w}$ and $\varphi_{\rm wg}$ are small. In practice, this limits the total number of photons present in the waveguide. In particular, at the next subleading order, there is a term $\sim\varphi^4$, which gives rise to a resonant Kerr effect that cannot be eliminated and leads to large, photon-number-dependent frequency shifts.  This problem can be mitigated by using an array of $N_{\rm c}$ identical couplers in series. The interaction then changes from $V(\varphi_{\rm wg},\varphi_{\rm w},\varphi_t)$ to $N_{\rm c}V(\varphi_{\rm wg}/N_{\rm c},\varphi_{\rm w}/N_{\rm c},\varphi_t)$, assuming an equal distribution of the flux over all SNAIL elements. Importantly, the flux drive $\varphi_t$ is applied to every loop separately and therefore does not scale as $\propto N_{\rm c}^{-1}$. The resulting expanded potential is
\begin{align}
V(\varphi)=\;&\mathrm{const.}+\frac{1}{54N_{\rm c}^2}E_J\varphi_t\varphi^3+\frac{1}{81N_{\rm c}^3}E_J\varphi^4\notag\\
&+\mathcal{O}(\varphi_t^2\varphi^2,\varphi_t^3\varphi,\varphi_t\varphi^5,\varphi^6,\dots).
\end{align}
We see that in this multi-coupler configuration, the term $\sim \varphi^4$ can be systematically suppressed compared to the interaction of interest, by increasing $N_{\rm c}$. 

For this general setup, the main conditions for the validity of the cooling master equation are as follows. First, the phase fluctuations of the waveguide and the waste mode must be smaller than unity: $\varphi^0_{\rm w}\sqrt{n_{\rm w}}/N_{\rm c},\varphi^0_1\sqrt{N_{\rm ph}}/N_{\rm c}<1$. Second, the Kerr and Lamb shifts that arise from the coupler detune the modes from the ideal equidistant spacing and $\omega_k= k\Delta +\delta_k$. The resulting frequency differences $\Omega_k=\delta_{k+1}-\delta_k$ must be smaller than the decay rate $\Gamma$, in order for the engineered cooling to be efficient. For the $k$-th mode, the total frequency shift can be estimated by
\begin{equation}
    %\hbar\widetilde{\omega}_k
    \hbar \delta_k =K_k^{\rm se}(n_k)+\sum_{k'}K_{k,k'}^{\rm cr}(n_{k'})+K_{k,\rm w}^{\rm cr}(n_{\rm w})+L_k.
\end{equation}
The self-Kerr term reads $K_1^{\rm se}(n_1)=6 E_J(\varphi_1^{0})^4n_1/(81 N_{\rm c}^3)$ for the fundamental mode and can be neglected for all other modes since their occupation numbers are typically much lower. The cross-Kerr term between  the fundamental mode and the $k$-th waveguide mode is related to the self-Kerr term by $K_{k,1}^{\rm cr}(n_1)=4K^{\rm se}_1(n_1)/k$ and the cross-Kerr term between the fundamental mode and the waste mode reads $K_{k,\rm w}^{\rm cr}(n_{\rm w})=24E_J(\varphi_k^0)^2(\varphi_{\rm w}^0)^2n_{\rm w}/(81 N_{\rm c}^3)$. Again, all other cross-Kerr terms can be neglected compared to those dominate terms that involve the condensate mode. Additionally, from the subleading interaction term $\varphi_t^2\varphi^2$ a Lamb shift of $L_k=6E_J\epsilon^2(\varphi_k^0)^2/(81 N_{\rm c})$ arises for each mode.

\rev{The shifts $\delta_k$ are acceptable as long as the resonance condition $\omega_{k+1}-\omega_k=\omega_{\rm w}-\omega_{\rm d}$ is fulfilled up to the bandwidth of the waste mode}, i.e., $|\delta_{k}-\delta_{k+1}| <\Gamma$. At the same time we must ensure $\Gamma \gg \Delta$ to resolve the individual modes and $g^2/\Gamma\gg \gamma$ to obtain a sufficiently fast engineered cooling. The device parameters summarized in Table~\ref{TabPara} satisfy all those constraints for the example of $N_{\rm c}=3$, but other parameter choices are possible as well.

Finally, we consider capacitive couplings induced by the SNAIL, which we have neglected so far. All Josephson junctions have a parallel capacitance, which gives rise to a total parallel capacitance $C_{\rm c}$. Although we can neglect the effect of this capacitance on the mode decomposition, it induces a charge-charge coupling between the waste and the waveguide modes, 
\begin{equation}
\frac{C_{\rm c}}{CC_\mathrm{w}}q_\mathrm{w}q_{\rm wg},
\end{equation}
where $q_\mathrm{w}$ is the waste charge operator and $q_{\rm wg}=q(x=0)$ the charge density operator at the end of the waveguide. For $C_{\rm c}/(\ell C)\lesssim10^{-5}$ the resulting coupling is still weak compared to the detuning between the waste mode and the closest waveguide mode. But it can induce an off-resonant linear loss mechanism. For the parameters of interest, $|\Delta-\omega_{\rm w}|\sim 0.1-1$ GHz, we estimate that the resulting capacitive loss rate is in the order of a few Hz, and thus comparable or smaller than the Ohmic damping rate $\gamma$.

\nocite{apsrev42Control}
\bibliographystyle{apsrev4-2}



\end{document}